# Design and Analysis of Novel Kernel Measure for Software Fault Localization

Vangipuram RadhaKrishna
Faculty of Information Technology
VNR Vignana Jyothi Institute of Engineering and Technology
Hyderabad, INDIA
vrkrishna2014@gmail.com

## ABSTRACT
The problem of software fault localization may be viewed as an approach for finding hidden faults or bugs in the existing program codes which are syntactically correct and give fault free output for some input instances but fail for all other input instances. Some of the reasons include logical errors, wrong interpretation of specification, coding errors. Finding such faults is not possible sometimes with the help of compilers. This is where the necessity and significance of software fault localization stems out. The main contribution for this work is to first introduce the block hit-miss function which relates block vectors of execution sequences of software code over sample runs performed and the decision vector which denotes fault or error free output. The similarity measure is applied to the block vector and decision vectors as input and the pair with maximum similarity is considered as faulty block.

## Categories and Subject Descriptors
D.2.5 **[Software Engineering]:** Testing and Debugging – Debugging Aids

## 1. INTRODUCTION
Exhaustive testing is an approach where we test the software developed using each possible test case. However, Exhaustive testing is not possible practically due to the number of combinations of test case input generated and is a typical NP-Hard Problem. In general, 100% testing is never possible. This is where a fault creeps in and then coins out suddenly which causes software failure. So, usually, Software testing concentrates on Effective Testing where the product code or software code is tested using various approaches of white box testing. The software reliability may be improved by performing effective testing and extensive debugging which is contradictory to the market conditions. When the bugs are noticed in testing phase only those bugs which have the highest priority and need to be removed immediately are considered to be significant and important.

Typically, in testing phase more and more bugs are released than those which can be debugged and hence the process of debugging turns out as bottleneck when we aim to achieve software reliability. We first define fault, error and failure. A fault also called as a bug or defect is the main cause of error coined out in the software system. Formally, we may define an Error as a state of the system which may cause software failure. A software failure is the event which essentially occurs as a result of observed output deviating from the expected output.

The objective of this research is to identify fault locations in the software code by defining a similarity measure and using this as part of clustering process so as to find the software faults in the program code. The approach we follow is to first find the block hits when each input instance is run on the program code. Then from this sample runs we record each input instance which gives correct result and wrong results or errors. An Error vector is formed using these input-output instances. A block hit means program control entering a particular block and performing some computations or scan of a block.

In our context we define clustering as an unsupervised learning process used to explore the underlying hidden structure of the faulty software code. In the context of software fault localization, we can map this problem to find the similarity between program block causing error and error vector generated for a set of legitimate input instances recorded against a set of legitimate expected output instances. As the input to the clustering algorithm is a usually a vector and also the similarity measure which we define in this paper uses the concept of feature vector, our first objective is to transform the program code into an equivalent vector format hence making it suitable for the design and application of the proposed similarity measure for fault localization.

In this research, the objective is to present the novel similarity measure which computes similarity score between each software block against the error vector generated for set of sample runs. We call this measure as SMSFL measure. The proposed measure is validated and analyzed for all the possible cases considering worst-case, best case, average cases.

We outline some related works in Section 2 and introduce the proposed similarity measure in section 3. Section 4 gives a simple case study to find faulty blocks. We finally conclude the paper in Section 5.

## 2. RELATED WORKS

Early software fault detection is very much important in software development process. If the fault or bug is left undetected in the initial stages of software development cycle, then it shall propagate to the next stages of the development cycle which is a serious hazard.

In [1], a detailed review of software fault localization techniques is discussed. Their work emphasizes the need for test suites software fault localization. Many times during software development process, most of the valuable resources, time is killed in debugging. The approach of feature selection for software fault localization is studied in [2]. Some of the works in software fault localization include [3, 4, 6, and 8]. Even after carrying out extensive testing, still defects remain un explored many times in the software development. This may be because of many reasons.

The concept of genetic invariants [5] is used to detect fault in a software program code which used automated approach. In [7, 9, 12, and 13], the software fault localization using similarity measure based approach is discussed. In [10], an equivalence based approach for software fault localization is addressed. An approach for behavioral analyzing of graphs is discussed in [11] to address fault localization.

In [14], the authors introduce the concept of hybrid fault localization. An approach for fault localization called software fault localization using the concept of cause tree is proposed in [15]. Their approach uses cause tree and organizes all the potential causes into a tree like structure.

## 3. PROPOSED APPROACH
### 3.1 Software Fault Localization

We define the proposed measure based on the fact that the fault occurring within a basic block of program code is based on the property of block hit and the block miss. Also, the hit or miss value of the block distribution has a significant contribution in evaluating and finding the fault by computing the similarity score value for the two vectors namely block vector and error vector.

The basic idea behind fault localization is to first form a matrix of order S X B where S is the number of input instances and B is the number of blocks in a program code. For each input instance if an error occurs we record it as 1 and if there is no error we record it as a 0. The basic blocks may be identified by following the procedure available in literature and also using principles of compiler design. We now introduce the ternary block-hit function which is then followed by definition of the proposed similarity measure for software fault detection.

We choose to represent the block-hit function by the notation, $H_c< B_{ik}, B_{jk} >$. The block-hit function is defined as shown in the Table.1.

Table.1 Block Hit Function

| $B_{ik}$ | $B_{jk}$ | $H_c< B_{ik}, B_{jk} >$ |
|---|---|---|
| Block miss (0) | Block miss (0) | -1 |
| Block miss (0) | Block hit (1) | 0 |
| Block hit (1) | Block miss (0) | 0 |
| Block hit (1) | Block hit (1) | 1 |

Here, $B_{ik}$, $B_{jk}$ are binary variables indicating the block hit or block miss for input instances $S_i$ and $S_j$ respectively with i and j being index of input instances and k representing block number.

The block hit is denoted by 1 and the block miss is denoted by 0. The Hit function denoted as $H_c<B_{ik}, B_{jk}>$ evaluates to any one of the three values 0, 1 or -1. The value 1 indicates the $k^{th}$ block is hit in both input instances and value of -1 indicates that the $k^{th}$ block is missed when run of both input instances, a value of 0 indicates that the combination of block hit-miss or block miss-hit.

Let $S_i$ and $S_j$ be the two input instances with the number of blocks in a given program code equal to 'm'. Let B be the block set representing 'm' blocks of the program code. We represent the two input instance vectors $S_i$ and $S_j$ as equivalent to the two binary sets of the form

$S_i = \{ B_{i1}, B_{i2}, B_{i3}, ... B_{im} \}$ and
$S_j = \{ B_{j1}, B_{j2}, B_{j3}, ... B_{jm} \}$

The notation $B_{im}$ or $B_{jm}$ corresponds to the block hit or block miss w.r.t $m^{th}$ block. The $m^{th}$ block hit corresponding to input sequence 'i' is denoted by $B_{im}$, and represented using a value 1 if the block is hit. i.e the execution enters the corresponding block. Alternatively, a block miss is indicated by a value equal to 0.

Once we obtain the input instance vectors, we must compute block feature vector which is a function of the block hit function defined in table.1.

Let $I_1$ and $I_2$ be any two run instances, when program code is run over sample inputs, then the block feature vector for two input instances is denoted as $BFV_{12}$, Formally represented as Block Feature-vector $[I_1, I_2]$,

$$BFV_{12} = [H_c< B_{11}, B_{21} >, H_c < B_{12}, B_{22} > .... H_c<B_{1m}, B_{2m} >] \quad (1)$$

Formally, the feature vector for any two input instances $I_i$ and $I_j$ with indices i and j may be defined mathematically as Block Feature-vector $[I_i, I_j]$,

$$BFV_{ij} = [H_c< B_{i1}, B_{j1} >, H_c < B_{i2}, B_{j2} > .... H_c<B_{im}, B_{jm} >] \quad (2)$$

The kernel function is denoted by SFLM and defined by eq.3 given below

$$SFLM = 0.5 * [1 + \frac{\sum_{k=1}^{k=m} \eta(B_{ik}, B_{jk})}{\sum_{k=1}^{k=m} \varphi(B_{ik}, B_{jk})}] \quad (3)$$

Where

$$\eta = \begin{cases} e^{-[1-h_c< B_{ik}, B_{jk}>]} & ; H_c < B_{ik}, B_{jk} > = 1 \\ 0 & ; H_c < B_{ik}, B_{jk} > = -1 \\ -e^{-h_c< B_{ik}, B_{jk}>} & ; H_c < B_{ik}, B_{jk} > = 0 \end{cases} \quad (4)$$

$$\varphi = \begin{cases} 0 ; H_C < B_{ik}, B_{jk} > \\ 1 ; H_C < B_{ik}, B_{jk} > \end{cases} \quad (5)$$

The value of the kernel function lies between 0 and 1. A value of 0 indicates that the similarity is minimum and a value of 1 indicates the similarity is maximum. The higher the similarity, the corresponding block has a chance of becoming faultier. i.e the block which has maximum similarity value is considered as the faulty block. From, this information we may perform, necessary changes to the program code so that it may transformed to fault free program code.

## 3.2 Algorithm for Software Fault Localization Using Proposed Kernel Measure

**Algorithm : Software Fault Localization**

**Software Fault Localization (Block vectors, Decision Vector)**

// Input : Block Vectors, Decision Class vector
// Output : Faulty Blocks

**Begin**
**Step-1:**
Obtain the program flow graph for the program code. The program flow graph may be obtained using the principles of compiler design.

**Step-2:**
From program flow graph, identify the program blocks. This may be done by finding the leader statements.

**Step-3:**
Create random sample input instances, by considering worst, average and best case situations.

For example, for sorting, the best case input is 1, 2, 3, 4, 5 when inputs are already sorted in increasing order. Alternately, the worst case input is 5, 4, 3, 2, 1 where all the elements are in non-increasing order. This is for the situation, when we are required to arrange elements in ascending order.

**Step-4:**
Run the program code on these legitimate input instances. Record corresponding outputs. During this execution process, record for each input instance, the corresponding block hits and block miss.

**Step-5:**
Classify the outputs as faulty or error-free.

**Step-6**:
Form a table with first 'm' columns indicating blocks and last column, denoting decision class. From step-5 and step-6, obtain the block vectors, denoted by B and decision vector, D.

**Step-7:**
Find similarity between the decision vector, D and the block vectors, B generated. The block vector which has the maximum similarity w.r.t decision vector is the faulty block.

## 4. CASE STUDY
Consider the following program code which is faulty c function for sorting rational numbers. This example is taken from [13]

Void Sort(int number, int *Numerator, int *Denominator)

{ /* block 1 */

```
int I , J, Temp;
for ( I= number-1;  I >=0 ; i-- )
{
/* block 2 */
for ( J=0; J < I; J++)
{
/* block 3 */
if (RationalGT(Numerator [J], Denominator [J],
Numerator[J+1], Denominator[J+1]))
{
/* block 4 */
Temp = Numerator [J];
Numerator[J] = Numerator [J+1];
Numerator[J+1] = Temp;
}
}
}
}
```

Consider two sequences as shown below

$$S1 = \{ \frac{1}{6} ; \frac{1}{5} ; \frac{1}{4} ; \frac{1}{2} \}$$

$$S2 = \{ \frac{3}{1} ; \frac{2}{2} ; \frac{4}{3} ; \frac{1}{4} \}$$

When the above sequences are input to the sort procedure, the first sequence does well, since it is sorted already. However, the second sequence leads to failure and the sort procedure does not perform as expected on this sequence. The output is faulty.

When, the above sort procedures is run on these two input sequences, the corresponding block hits and miss is as depicted in Table.2. The symbol ✓ in the last column indicates that an error has occurred and the symbol ✗ denotes no error. Similarly, for all other columns, the symbol ✓ indicates that execution enters the corresponding block and the symbol ✗ denotes no entry into the block.

Table.2 Sample Run

| Input | Blocks | | | | | | Decision |
|---|---|---|---|---|---|---|---|
| Input Sequence | 0 | 1 | 2 | 3 | 4 | 5 | Error |
| S1 | ✓ | ✓ | ✓ | ✓ | ✗ | ✓ | ✗ |
| S2 | ✓ | ✓ | ✓ | ✓ | ✓ | ✓ | ✓ |

In Table.2, the execution enters block-4 for the input sequence S1, while it is a miss when input sequence S2 is specified. From Table.2, the block column vectors and decision column vector generated may be represented as shown below

$$c0 = \begin{bmatrix} 1 \\ 1 \end{bmatrix}$$
$$c1 = \begin{bmatrix} 1 \\ 1 \end{bmatrix}$$
$$c2 = \begin{bmatrix} 1 \\ 1 \end{bmatrix}$$
$$c3 = \begin{bmatrix} 1 \\ 1 \end{bmatrix}$$
$$c4 = \begin{bmatrix} 0 \\ 1 \end{bmatrix}$$
$$c5 = \begin{bmatrix} 1 \\ 1 \end{bmatrix}$$
$$D = \begin{bmatrix} 0 \\ 1 \end{bmatrix}$$

Once the column vectors are obtained, transpose the column vectors and decision vector to get the corresponding row vectors. This is as shown below using R0 to R5 and decision row vector denoted by D.

$$R0 = [1 \quad 1]$$
$$R1 = [1 \quad 1]$$
$$R2 = [1 \quad 1]$$
$$R3 = [1 \quad 1]$$
$$R4 = [0 \quad 1]$$
$$R5 = [1 \quad 1]$$
$$D = [0 \quad 1]$$

Now, find the similarity between each row vector and the decision column using the proposed measure. The row vector which has the maximum similarity with the Decision vector is considered as faulty block.

In our case, the block 4 is faulty block, having similarity value as 1 while other row vectors have similarity as 0.5. Hence, we must eliminate the fault or error w.r.t block 4, so that the elements may be sorted as required.

The approach of finding software fault localization using proposed measure is simple and effective and may be applied to complex program modules to locate, logical software faults if any. The measure may also be used for achieving software reuse by making necessary changes and applying suitably.

## 5. CONCLUSIONS

Software development process is not free from faults, errors and failures. If a fault exists and goes unnoticed, then this affects the subsequent stages of the development process. Also, most of the software development time gets killed in the debugging phase. The problem of software fault localization using the concept of similarity measure is discussed in this paper. By localizing faults, which occur early in the software development process, we may reduce the overall time and cost. The idea of using similarity measure in this work is mainly to locate the blocks of code which are logically wrong and then output those blocks of code which leads to software failure with the existence of faults. The proposed measure takes as input, the sequences which are functions of program blocks and outputs the list of all faulty blocks. In future, we may extend this to recover software architectures, lines of code etc.